\pdfoutput=1

\documentclass[11pt, letter,fleqn]{article}
\usepackage{graphics,color}
\usepackage{caption}
\usepackage{subcaption}
\usepackage{cprotect}
\usepackage{epstopdf}
\usepackage{graphicx}
\usepackage{verbatim}
\usepackage{float}
\usepackage{listings}
\usepackage{tikz}
\usepackage[section]{placeins}
\usepackage{multirow}
\usepackage{pdflscape}	
\usepackage{enumitem}  
\usepackage{cite}

\usepackage{refstyle}
\usepackage{amsmath, amssymb,amsthm}
\usepackage{chngcntr}
\usepackage[toc,page]{appendix}
\counterwithin{equation}{section}

\title{On the Hill's Spherical Vortex in Fluid and Plasma, its Generalization, and Stability}

\author{  Jason M. Keller\footnotemark[1], ~~Alexei
  F. Cheviakov\footnotemark[2] \\ {\small
    \emph{Department of Mathematics and Statistics, University of
      Saskatchewan, Saskatoon, S7N 5E6 Canada}}
}

 {\theoremstyle{definition}

{\theoremstyle{definition} }
{\theoremstyle{definition} }

\def\const{\hbox{\rm const}}

\def\grad{\mathop{\hbox{\rm grad}}}

\def\div{\mathop{\hbox{\rm div}}}
\def\curl{\mathop{\hbox{\rm curl}}}
\def\vec#1{{\boldsymbol{\rm #1}}}  

\def\beq{\begin{equation}}
\def\eeq{\end{equation}}
\def\barr{\begin{array}{ll}}
\def\earr{\end{array}}

\begin{document}


\maketitle

\begin{abstract}

In 1894 M.J.M. Hill published an article describing a spherical vortex moving through a stationary fluid. Using cylindrical coordinates and assuming the azimuthal velocity component zero, Hill found a simple solution that described this flow. A similar modern problem in the MHD framework was put forth in 1987 by A. A. Bobnev and in 1995 by R. Kaiser and D. Lortz who applied the setup to model a ball lighting. We present a much simpler derivation of Hill's spherical vortex using the Bragg-Hawthorne equation. In particular, by using the moving frame of reference, the Euler equations reduce to equilibrium flow which are equivalent to the static equilibrium MHD equations up to relabelling. A new generalized version of Hill's spherical vortex with a nonzero azimuthal component is derived. A physical solution to the static equilibrium MHD equations is computed by looking at a separated solution to the Grad-Shafranov equation in spherical coordinates. Finally, the stability of Hill's spherical vortex is examined by performing an axisymmetric perturbation described; it is shown that the Hill's spherical vortex is linearly unstable with respect to certain kinds of small perturbations.

\end{abstract}

\section{Introduction}
In 1894 Micaiah John Muller Hill published an article describing a sphere moving symmetrically with regards to an axis through a stationary fluid. Using cylinderical coordinates and assuming that the azimuthal velocity component is zero, Hill was able to find a simple solution that describes this fluid flow. This solution and the method that it was computed is available in \cite{hill1894vi}. A similar modern problem in the MHD framework was put forth in 1987 by A. A. Bobnev in which they considered a spherical vortex moving in an ideally conducting fluid \cite{Bobnev}. In this work, several small mistakes were made. Interestingly enough, in 1995 R. Kaiser and D. Lortz again considered the problem of a spherical vortex in MHD equilibrium to model ball lighting \cite{kaiser1995ball} essentially re-deriving the solution A. Bobnev found in \cite{Bobnev}. In the following chapter, a modern and much simpler derivation of Hill's spherical vortex using the Bragg-Hawthorne equation, (which was first derived in 1898 by William Mitchinson Hicks and only gained popularity after being re-derived in 1950 by William Hawthorne and Stephen Bragg) will be shown to emphasize the usefulness of the Bragg-Hawthorne equation for such problems. By using the moving frame of reference the Euler equations reduce to equilibrium flow which are equivalent to the static equilibrium MHD equations up to relabelling. Next, the spherical vortex in an ideally conducting fluid is computed similar to methods in both \cite{Bobnev} and \cite{kaiser1995ball}. Using results from the previous two sections, a new generalized version of Hill's spherical vortex is put forth. After this, a physical solution to the static equilibrium MHD equations is computed by looking at a separated solution to the Grad-Shafranov equation in spherical coordinates and lastly, the stability of Hill's spherical vortex is examined by performing an axisymmetric perturbation described in \cite{pozrikidis1986nonlinear}. A similar analysis for the new generalized Hill's spherical vortex is also attempted.

\section{Hill's spherical vortex: a modern derivation}

A sphere of radius $R$ moving through a stationary fluid directed along the $z$ axis can be modelled with the incompressible Euler equations. Starting with the equations of motion for an incompressible fluid

\begin{subequations}\label{eq:Euler3}
	\begin{equation}
		\frac{\partial \vec{V}}{\partial t} + (\vec{V} \cdot \nabla) \vec{V} = -\frac{1}{\rho}\grad P,
	\end{equation}
	\begin{equation}
		\div {\vec{V}} = 0,
	\end{equation}
\end{subequations}
the well known result that the incompressible Euler equations are invariant under a general Galilean transformations motivates the following change of variables

\begin{equation}\label{eq:Gal}
	\vec{V}(\vec{r},t) = \vec{\tilde{v}}(\vec{{{r}}} - Z(t)\vec{e}_z) + Z'(t)\vec{e}_z, \quad P(\vec{r},t) = \tilde{P}(\vec{r} - Z(t)\vec{e}_z).
\end{equation}
Here $Z(t)$ is an arbitrary function of time and $\tilde{\vec{v}}$, $\tilde{P}$ denote fluid parameters measured in the corresponding moving frame of reference.
\medskip

Assuming that the moving frame of reference is moving at the same speed as the spherical vortex, and the density is constant, after omitting the tilde on the new variables, the Euler equations can be written as

\begin{subequations}\label{eq:Euler_eq3}
	\begin{equation}
		\curl \vec{v} \times \vec{v} = \grad H,
	\end{equation}
	\begin{equation}
		\div \vec{v} = 0,
	\end{equation}
\end{subequations}
where
\begin{equation}
	H = -\left(\frac{P}{\rho} + \frac{1}{2} |\vec{v}|^2\right)
\end{equation}
is a modified pressure term. In the rest of this section $H$ will simply be refereed to as the pressure. As one can see, Assuming that the motion is axially symmetric it is natural to use cylindrical coordinates and set $\vec{v}$ and $H$ independent of $\phi$. In doing so, one can reduce (\ref{eq:Euler_eq3}) to the well known Bragg-Hawthorne equation

\begin{equation}\label{eq:B_Heq}
	\frac{\partial^2 \psi}{\partial r^2} + \frac{\partial^2 \psi}{\partial z^2} - \frac{1}{r}\frac{\partial \psi}{\partial r} + F(\psi)F'(\psi) = r^2 H'(\psi),
\end{equation}
where
\begin{equation}
	\vec{v} =  \frac{\psi_z}{r} \vec{e}_r + \frac{F(\psi)}{r}\vec{e}_\phi + \frac{-\psi_r}{r} \vec{e}_z,
\end{equation}
and $F$, $H$ are arbitrary functions of $\psi$, where $\psi$ is the stream function discussed in Section 6.2. Chapter 1. Following Hill's assumption who considered a two-component axially symmetric flow, the azimuthal component of the velocity is set to zero, giving the condition

\begin{equation}\label{eq:F_cond}
	F(\psi) = 0.
\end{equation}

From this, the vorticity becomes
\begin{equation}
	\vec{\omega} = r^2H'(\psi)\vec{e}_{\phi}.
\end{equation}

Note that when the pressure is constant: $H = H_0$, one has $\vec{\omega} = 0$, which corresponds to an irrotational flow, (\ref{eq:F_cond}) also gives a simplified Bragg-Hawthorne equation

\begin{equation}\label{eq:B_Heq_simp}
	\frac{\partial^2 \psi}{\partial r^2} + \frac{\partial^2 \psi}{\partial z^2} - \frac{1}{r}\frac{\partial \psi}{\partial r} = -r^2 H'(\psi),
\end{equation}
with
\begin{equation}\label{vel_comp}
	\vec{v} =  \frac{\psi_z}{r} \vec{e}_r  + \frac{-\psi_r}{r} \vec{e}_z.
\end{equation}

The arbitrary function is chosen as the highest power series expansion in $\psi$ such that the (\ref{eq:B_Heq_simp}) becomes separable in spherical coordinates and the asymptotics of the pressure $H(\psi)$ behaves properly. As far as separability of (\ref{eq:B_Heq_simp}) goes, $H(\psi)$ cannot be of higher degree then linear in $\psi$. In regards to the asymptotics, the pressure far away from the sphere must not change and needs to be the ambient pressure $H_0$. This gives the best choice for $H(\psi)$ to be broken into two pieces that match at the boundary

\begin{equation}\label{eq:piecepress}
	H(\psi) =
	\begin{cases}
		H_0 - 10\delta \psi, & \rho < R\\
		
		H_0 . & \rho > R
	\end{cases}
\end{equation}
Here the coefficient $10\delta$ is only chosen in this way to make the calculation cleaner. The problem is now be decomposed into two pieces: the rotational flow inside of the sphere with pressure linear in $\psi$, and the irrotational flow outside of the sphere with constant pressure. \medskip

\begin{enumerate}
	\item{Rotational flow inside the sphere}
	\begin{subequations}
		\begin{equation}
			H(\psi) = H_0 - 10\delta \psi
		\end{equation}
		\begin{equation}\label{eq:B_Heq_lin_inside}
			\frac{\partial^2 \psi}{\partial r^2} + \frac{\partial^2 \psi}{\partial z^2} - \frac{1}{r}\frac{\partial \psi}{\partial r} = 10\delta r^2.
		\end{equation}
	\end{subequations}
	
	\item{Irrotational flow outside the sphere}
	\begin{subequations}
		\begin{equation}
			H(\tilde{\psi}) = H_0
		\end{equation}
		\begin{equation}\label{eq:B_Heq_lin_outside}
			\frac{\partial^2 \tilde{\psi}}{\partial r^2} + \frac{\partial^2 \tilde{\psi}}{\partial z^2} - \frac{1}{r}\frac{\partial \tilde{\psi}}{\partial r} = 0.
		\end{equation}
	\end{subequations}
\end{enumerate}

Along with these two equations, there is the condition that both pieces must have matching pressure and velocity components at the boundary of the sphere ($r^2 + z^2 = R^2$). For matching pressure, this implies that for the inside solution, $\psi(r,z) = 0$ when $r^2 + z^2 = R^2$. It turns out that one can effectively seek solutions to (\ref{eq:B_Heq_lin_inside}) and (\ref{eq:B_Heq_lin_outside}) in spherical coordinates, in the separated form $\psi(\rho,\theta) = R(\rho) \Theta(\theta)$. Here standard spherical coordinates are related to cylindrical coordinates by $r = \rho \sin \theta$, $z = \rho \cos \theta$. Converting the above problem into spherical coordinates gives

\begin{enumerate}
	\item{Rotational flow inside the sphere}
	\begin{subequations}\label{eq:inside11}
		\begin{equation}
			H(\psi) = H_0 - 10\delta \psi
		\end{equation}
		\begin{equation}\label{eq:Hill_inside}
			\left[ \frac{\partial^2}{\partial \rho^2} + \frac{\sin \theta}{\rho^2} \frac{ \partial}{\partial \theta} \left(\frac{1}{\sin \theta} \frac{\partial}{\partial \theta}\right)\right] \psi = 10 \delta \rho^2 \sin^2{\theta}.
		\end{equation}
	\end{subequations}
	
	\item{Irrotational flow outside the sphere}
	\begin{subequations}\label{eq:inside22}
		\begin{equation}
			H(\tilde{\psi}) = H_0
		\end{equation}
		\begin{equation}
			\left[ \frac{\partial^2}{\partial \rho^2} + \frac{\sin \theta}{\rho^2} \frac{ \partial}{\partial \theta} \left(\frac{1}{\sin \theta} \frac{\partial}{\partial \theta}\right)\right] \psi = 0.
		\end{equation}
	\end{subequations}
\end{enumerate}

The velocity components inside and outside are given by
\begin{equation}
	\vec{v}_{in} =  \frac{1}{\rho^2\sin \theta}\frac{\partial \psi}{\partial \theta} \vec{e}_{\rho} - \frac{1}{\rho \sin \theta}\frac{\partial \psi}{\partial \rho} \vec{e}_{\theta},
\end{equation}
and
\begin{equation}
	\vec{v}_{out} =  \frac{1}{\rho^2\sin \theta}\frac{\partial \tilde{\psi}}{\partial \theta} \vec{e}_{\rho} - \frac{1}{\rho \sin \theta}\frac{\partial \tilde{\psi}}{\partial \rho} \vec{e}_{\theta}.
\end{equation}

respectively. Along with this, the matching conditions and the need for $\psi(\rho,\theta)$ to be regular at $\rho = 0$ give the following four boundary conditions \begin{equation}\label{eq:matching}
	\psi(R,\theta) = 0, \quad |\psi(0,\theta)| < \infty,  \quad \frac{\partial \psi}{\partial \theta} \bigg|_{\rho = R} = \frac{\partial \tilde{\psi}}{\partial \theta} \bigg|_{\rho = R}, \quad \frac{\partial \psi}{\partial \rho} \bigg|_{\rho = R} = \frac{\partial \tilde{\psi}}{\partial \rho} \bigg|_{\rho = R}.
\end{equation}

A general solution for the inhomogeneous inside equation  (\ref{eq:Hill_inside}) is sought in the form of $\psi(\rho,\theta) = \psi(\rho,\theta)_{gen} + \psi(\rho,\theta)_{part}$ where $\psi(\rho,\theta)_{gen}$ is a general solution to the homogeneous version of (\ref{eq:Hill_inside}) given by
\begin{equation}\label{eq:hillhom}
	\left[ \frac{\partial^2}{\partial \rho^2} + \frac{\sin \theta}{\rho^2} \frac{ \partial}{\partial \theta} \left(\frac{1}{\sin \theta} \frac{\partial}{\partial \theta}\right)\right] \psi = 0.
\end{equation}
and $\psi(\rho,\theta)_{part}$ is a particular solution to (\ref{eq:Hill_inside}). A particular solution is found to be
\begin{equation}
	\psi(\rho,\theta)_{part} = \delta\rho^4\sin^2\theta. 	
\end{equation}
The general solution to (\ref{eq:hillhom}) is obtained by a separated solution $\psi(\rho,\theta) = R(\rho) \Theta(\theta)$. Upon substituting the separated form into (\ref{eq:hillhom}) one arrives at the two ODEs

\begin{equation}\label{eq:rho_ode1}
	\rho^2 R'' - \mathcal{C}R = 0,
\end{equation}

\begin{equation}\label{eq:theta_ode1}
	\big((-\csc \theta) \Theta'\big)' = \mathcal{C} (\csc \theta) \Theta,
\end{equation}
where $\mathcal{C}$ is a separation constant to be determined. Using the change of variables

\begin{equation}
	t = \cos\theta, \quad \Theta(\theta) = T(t),
\end{equation}
the equation (\ref{eq:theta_ode1}) becomes
\begin{equation}\label{eq:Tequation1}
	(1-t^2)T''(t) + \mathcal{C}T(t) = 0.
\end{equation}
This ODE can be related to the associated Legendre ODE with the transformation
\begin{equation}
	T(t) = \sqrt{1 - t^2}P(t)
\end{equation}
leading to
\begin{equation}\label{eq:Legendre_m11}
	(1 - t^2)P''(t) - 2tP'(t) + \left(\mathcal{C} - \frac{1}{1-t^2}\right)P(t) = 0.
\end{equation}
The equation (\ref{eq:Legendre_m11}) is related to the associated Legendre ODE \cite{kaiser1995ball}.
\begin{equation}\label{eq:Legendre1}
	(1 - x^2)\tilde{P}''(x) - 2x\tilde{P}'(x) + \left(l(l+1) - \frac{m^2}{1 - x^2}\right)\tilde{P}(x) = 0
\end{equation}
Clearly (\ref{eq:Legendre_m11}) is the same as (\ref{eq:Legendre1}) when $m = 1$ and $\mathcal{C} = l(l+1)$. The equation (\ref{eq:Legendre1}) has nonsingular solutions on the interval $[-1,1]$ only when $l$ and $m$ are integer values \cite{arfken1999mathematical}. For $m = 1$, the associated Legendre polynomials have the form

\begin{equation}
	P_l(x) = -\sqrt{1 - x^2}\frac{d}{dx}\mathcal{P}_l(x),
\end{equation}
where $\mathcal{P}_l$ refers to the lth order Legendre polynomial. One then arrives at the regular solutions to (\ref{eq:Tequation1})
\begin{equation}
	T_l(t) = -(1 - t^2)\frac{d}{dt}\mathcal{P}_l.
\end{equation}
which can be written as
\begin{equation}
	T_l(t) = (l+1)\mathcal{P}_{l+1}(t) - (l+1)t\mathcal{P}_l(t).
\end{equation}
This gives $\Theta(\theta)$ as
\begin{equation}\label{eq:thetasol}
	\Theta_l(\theta) = (l+1)\mathcal{P}_{l+1}(\cos\theta) - (l+1)\cos\theta \ \mathcal{P}_l(\cos\theta).
\end{equation}
The value $\mathcal{C} = l(l+1)$ can now be substituted into (\ref{eq:rho_ode1}) giving
\begin{equation}
	\rho^2 R''(\rho) - l(l+1)R(\rho) = 0.
\end{equation}
This has the solution
\begin{equation}
	R_l(\rho) = a_l\rho^{l+1} + b_l\rho^{-l}.
\end{equation}
As the solution is required to be regular at $\rho = 0$, $b_l$ will be set to zero. A separated solution to the homogeneous PDE (\ref{eq:hillhom}) is therefore
\begin{equation}\label{eq:psi_sep}
	\psi_l(\rho,\theta) = a_l\rho^{l+1}\Theta_l(\theta),
\end{equation}
giving the solution for $\psi$ inside of the sphere as
\begin{equation}
	\psi(\rho,\theta) = \delta\rho^4\sin^2\theta + \sum_{l = 0}^{\infty}a_l\rho^{l+1}\Theta_l(\theta).
\end{equation}
\medskip

Using the condition that the pressure must match at the boundary which reduces to the condition that $\psi(R,\theta) = 0$ as specified in (\ref{eq:matching}) gives
\begin{equation}\label{eq:cond}
	\sum_{l = 0}^{\infty}a_l R^{l+1}\Theta_l(\theta) = -\delta R^4\sin^2\theta.
\end{equation}
The solutions $\Theta_l(\theta)$ form a complete orthogonal basis as (\ref{eq:theta_ode1}) is a classical Sturm-Liouville second-order linear ODE with weight $w(\theta) = -\csc \theta$. Using the observation that $\Theta_1(\theta) = -\sin^2\theta$ equation (\ref{eq:cond}) can be written as
\begin{equation}
	\sum_{l = 0}^{\infty}a_l R^{l+1}\Theta_l(\theta) = \delta R^4\Theta_l(\theta).
\end{equation}
By multiplying the above equation by $-\csc \theta \Theta_l(\theta)$ and integrating from $0 < \theta < \pi$ one arrives at

\begin{equation}
	a_l R^{l+1} = \frac{-\int_0^\pi \delta R^4 \csc \theta \Theta_1(\theta) \Theta_l(\theta) d\theta}{-\int_0^\pi\csc \theta (\Theta_l (\theta))^2 d\theta}.
\end{equation}
The right hand side is zero due to the orthogonality of $\Theta_l(\theta)$ for all $l$ except when $l = 1$. In this case one obtains the condition that

\begin{equation}
	a_1 = \delta R^2.
\end{equation}

Therefore the solution inside the sphere can be written as
\begin{equation}
	\psi(\rho,\theta) =		\delta \rho^2\sin^2\theta(\rho^2 - R^2).
\end{equation}

For outside of the sphere, the solution is the same as the homogeneous solution to \ref{eq:Hill_inside} given by

\begin{equation}
	\tilde{\psi}(\rho,\theta) = \sum_{l = 0}^{\infty}\left(c_l\rho^{l+1} + \frac{d_1}{\rho}\right)\Theta_l(\theta).
\end{equation}

The forth condition in (\ref{eq:matching}) gives the condition that

\begin{equation}\label{eq:cond2}
	\sum_{l = 0}^{\infty}\left(c_l(l+1) R^{l} - \frac{d_1}{R^2}\right)\Theta_l(\theta) = 3\delta R^3\sin^2 \theta.
\end{equation}

Using the orthogonality of $\Theta_l(\theta)$ as discussed before, $l = 1$. Lastly, the third condition gives

\begin{equation}
	\left(c_lR^{2} + \frac{d_1}{R}\right) = 0.
\end{equation}

giving $c_1 = -d_1/R^3$. Substituting this back into ($\ref{eq:cond2})$ with $l = 1$ one achieves the complete solution

\begin{equation}\label{eq:sol1}
	\psi(\rho,\theta) =
	\begin{cases}
		\delta \rho^2\sin^2\theta(\rho^2 - R^2), & \rho < R\\
		
		\frac{2}{3} \delta R^2\sin^2\theta \left(\frac{\rho^3 - R^3}{\rho}\right). & \rho > R
	\end{cases}
\end{equation}
This can be written in cylindrical coordinates as

\begin{equation}
	\psi(r,z) =
	\begin{cases}
		\delta\left((r^2z^2 + r^4) -R^2r^2\right), & r^2 + z^2 < R^2\\
		
		\frac{2}{3}\delta R^2 r^2\left(1 - \frac{R^3}{(r^2 + z^2)^{3/2}}\right). & r^2 + z^2 > R^2
	\end{cases}.
\end{equation}

The velocity components can be computed from (\ref{vel_comp}) to be

\begin{equation}\label{eq:vel1}
	v_r =
	\begin{cases}
		2\delta rz, & r^2 + z^2 < R^2\\
		
		\frac{2\delta R^5rz}{(r^2 + z^2)^{5/2}}, & r^2 + z^2 > R^2
	\end{cases}
\end{equation}

\begin{equation}\label{eq:vel2}
	v_z =
	\begin{cases}
		2\delta\left(R^2 -r^2 - z^2\right), & r^2 + z^2 < R^2\\
		
		\frac{4}{3}\delta R^2 + \frac{2\delta R^5}{3}\frac{(r^2 - 2z^2)}{(r^2 + z^2)^{5/2}}, & r^2 + z^2 > R^2
	\end{cases}.
\end{equation}

Moving back into the lab frame with the transformation given by (\ref{eq:Gal}) one arrives at

\begin{equation}
	V_r =
	\begin{cases}
		2\delta r(z - Z(t)), & r^2 + (z-Z(t))^2 < R^2\\
		
		\frac{2\delta R^5r(z - Z(t))}{(r^2 + (z - Z(t))^2)^{5/2}}, & r^2 + (z - Z(t))^2 > R^2
	\end{cases}
\end{equation}

\begin{equation}
	V_z =
	\begin{cases}
		Z'(t) + 2\delta\left(R^2 -r^2 - (z - Z(t))^2\right), & r^2 + (z - Z(t))^2 < R^2\\
		
		Z'(t) + \frac{4}{3}\delta R^2 + \frac{2\delta R^5}{3}\frac{(r^2 - 2(z - Z(t))^2)}{(r^2 + (z - Z(t))^2)^{5/2}}. & r^2 + (z - Z(t))^2 > R^2
	\end{cases}
\end{equation}
The pressure in the stationary frame of reference is given by
\begin{equation}\label{eq:hillpressure}
	H(r,z) =
	\begin{cases}
		H_0 - 10\delta^2 \left(r^2\left((z-Z(t))^2 + r^2 - R^2)\right)\right), & r^2 + (z-Z(t))^2 < R^2\\
		
		H_0. & r^2 + (z-Z(t))^2 > R^2
	\end{cases}
\end{equation}

\begin{figure}[htb!]	
	\begin{center}
		\includegraphics[width = .7\textwidth]{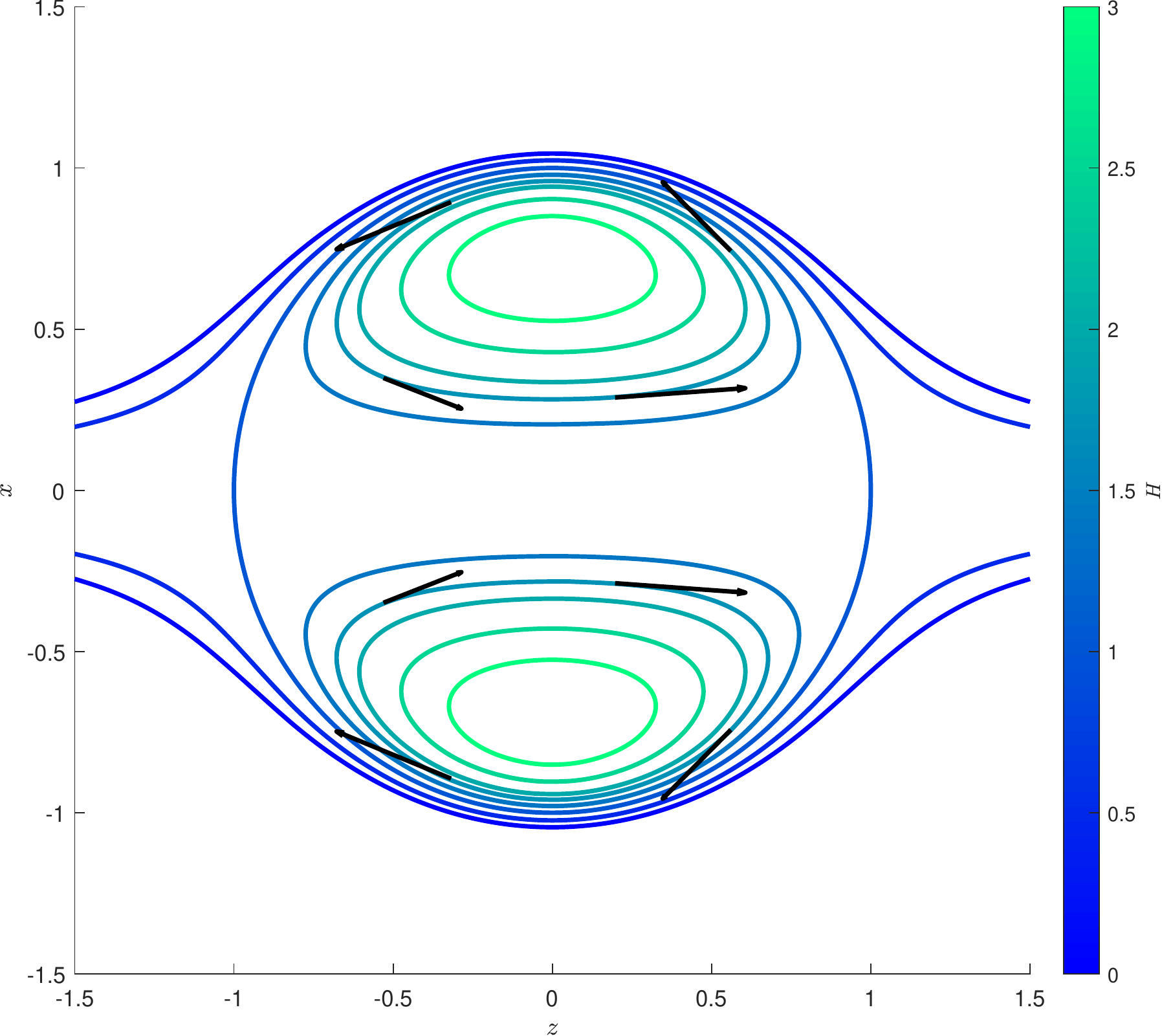}
	\end{center}
	\caption{\label{fig:Level_curves_of_xi}A cross-section of surfaces $H(\psi) = \const$ in the lab frame given by (\ref{eq:hillpressure1}). Here $R = 1$, $H_0 = 1$, $\delta = 1$ and $t = 0$. The black arrows correspond to the velocity vectors on a given surface. By the first equation of (\ref{eq:Euler_eq3}), both $\vec{v}$ and $\curl \vec{v}$ are tangent to this surface.}\label{fig:hills}
\end{figure}

One additional boundary condition that can be considered is the behaviour of the velocity far away from the spherical vortex. In particular, if the fluid that the sphere is moving through is stationary, it is natural to demand $v_r, v_z \to 0$ as $r^2 + z^2 \to \infty$. The first limit for $v_r$ is trivially satified

\begin{equation}
	\lim_{r^2 + z^2 \to \infty} v_r = 0.
\end{equation}
however, for the $z$ component of velocity, $v_z$ one gets
\begin{equation}
	\lim_{r^2 + z^2 \to \infty} v_z = Z'(t) + \frac{4}{3}\delta R^2 = 0.
\end{equation}
This gives the additional condition that $Z'(t) = -\frac{4}{3}\delta {R^2}$. This implies the interesting result that the group velocity of the moving spherical vortex is constant with a speed that is proportional to the square of the radius. In this case, the solution depending on the freedom of $R$ and $\delta$ can be written completely in terms of
\begin{equation}
	Z(t) = Z_0-\frac{4}{3}\delta {R^2}t
\end{equation}
as
\begin{equation}
	V_r =
	\begin{cases}
		2\delta r(z + \frac{4}{3}\delta {R^2}t), & r^2 + (z+\frac{4}{3}\delta {R^2}t)^2 < R^2\\
		
		\frac{2\delta R^5r(z+\frac{4}{3}\delta {R^2}t)}{(r^2 + (z+\frac{4}{3}\delta {R^2}t)^2)^{5/2}}, & r^2 + (z+\frac{4}{3}\delta {R^2}t)^2 > R^2
	\end{cases}
\end{equation}

\begin{equation}
	V_z =
	\begin{cases}
		-\frac{4}{3}\delta {R^2} + 2\delta\left(R^2 -r^2 - (z+\frac{4}{3}\delta {R^2}t)^2\right), & r^2 + (z+\frac{4}{3}\delta {R^2}t)^2 < R^2\\
		
		-\frac{4}{3}\delta {R^2} + \frac{4}{3}\delta R^2 + \frac{2\delta R^5}{3}\frac{(r^2 - 2(z+\frac{4}{3}\delta {R^2}t)^2)}{(r^2 + (z+\frac{4}{3}\delta {R^2}t)^2)^{5/2}}. & r^2 + z+\frac{4}{3}\delta {R^2}t)^2 > R^2
	\end{cases}
\end{equation}
with the pressure profile in the stationary frame as

\begin{equation}\label{eq:hillpressure1}
	H(r,z) =
	\begin{cases}
		H_0 + 10\delta^2 \left(r^2\left((z+\frac{4}{3}\delta {R^2}t)^2 + r^2 - R^2)\right)\right), & r^2 + (z+\frac{4}{3}\delta {R^2}t)^2 < R^2\\
		
		H_0. & r^2 + (z+\frac{4}{3}\delta {R^2}t)^2 > R^2
	\end{cases}
\end{equation}
Level curves of $H(r,z)$ can be seen in Figure \ref{fig:hills}.

\section{A stationary spherical MHD vortex}

A similar problem to Hill's spherical vortex is the concept of a spherical vortex moving through an ideally conducting fluid. With this problem, negligibly small fluid motion ($\vec{V} = 0$) is assumed which gives the starting point as the static equilibrium MHD equations
\begin{subequations}\label{eq:MHDst2}
	\begin{equation}
		\curl \vec{B} \times \vec{B}= \mu \grad P,
	\end{equation}
	\begin{equation}
		\div \vec{B} = 0.
	\end{equation}
\end{subequations}
Here, the main differences between Hill's spherical vortex and this stationary conducting spherical vortex is: the search for $\vec{v}$ inside and outside the sphere is replaced with the search for $\vec{B}$, and the azimuthal component of this magnetic field is \emph{not} assumed to be zero. Two assumptions of this conducting spherical vortex are: the pressure goes to a constant value taken to be zero at the boundary of the sphere (similar to Hill's spherical vortex), and every magnetic field component goes to zero at the boundary. The last condition here regarding the magnetic field is chosen in this way because the asymptotic behaviour of the magnetic field must decay at least as quickly as a dipole moment, but it was shown in \cite{kaiser1995ball}, that the only solution outside of the sphere consistent with the inside pressure and magnetic field that has the proper asymptotic behaviour is when $\vec{B} = 0$.
\medskip

The spherical vortex is assumed to have inherent axial symmetry which allows the reduction of (\ref{eq:MHDst2}) to the Grad-Shafranov equation

\begin{equation}\label{eq:G-Seq}
	\frac{\partial^2 \psi}{\partial r^2} + \frac{\partial^2 \psi}{\partial z^2} - \frac{1}{r}\frac{\partial \psi}{\partial r} + I(\psi)I'(\psi) = -r^2 P'(\psi).
\end{equation}
where the magnetic field components are given by
\begin{equation}\label{eq:B-comp}
	\vec{B} =  \frac{\psi_z}{r} \vec{e}_r + \frac{I(\psi)}{r}\vec{e}_\phi - \frac{\psi_r}{r} \vec{e}_z.
\end{equation}
Inside of the sphere, the pressure $P(\psi)$ and the arbitrary function related to the toroidal magnetic field $I(\psi)$ are taken to be linear (as any higher power series expansion of $P(\psi)$ and $I(\psi)$ makes (\ref{eq:G-Seq}) not separable in spherical coordinates). Therefore, these arbitrary functions are written as
\begin{equation}\label{eq:mhdvortfuncs}
	P(\psi) = P_0 - \gamma \psi, \quad I(\psi) = \lambda \psi.
\end{equation}
The Grad-Shafranov equation now becomes a second order linear homogeneous PDE. This equation is now converted to spherical coordinates
\begin{equation}\label{eq:GS_spherical1}
	\left[ \frac{\partial^2}{\partial \rho^2} + \frac{\sin \theta}{\rho^2} \frac{ \partial}{\partial \theta} \left(\frac{1}{\sin \theta} \frac{\partial}{\partial \theta}\right) + \lambda^2\right] \psi = \gamma \rho^2 \sin^2{\theta},
\end{equation}
where the magnetic field is given by
\begin{equation}\label{eq:magn}
	\vec{B} =  \frac{1}{\rho^2\sin \theta}\frac{\partial \psi}{\partial \theta} \vec{e}_{\rho} + \frac{I(\psi)}{\rho \sin \theta}\vec{e}_\phi - \frac{1}{\rho \sin \theta}\frac{\partial \psi}{\partial \rho} \vec{e}_{\theta}.
\end{equation}

Following a similar method to the previous section, $\psi(\rho,\theta) = \psi(\rho,\theta)_{gen} + \psi(\rho,\theta)_{part}$ where $\psi(\rho,\theta)_{gen}$ is a general solution to the homogeneous version of (\ref{eq:GS_spherical1}) given by
\begin{equation}\label{eq:Bobhom}
	\left[ \frac{\partial^2}{\partial \rho^2} + \frac{\sin \theta}{\rho^2} \frac{ \partial}{\partial \theta} \left(\frac{1}{\sin \theta} \frac{\partial}{\partial \theta}\right) + \lambda^2\right] \psi = 0,
\end{equation}

A particular solution to (\ref{eq:GS_spherical1}) is found to be
\begin{equation}\label{eq:part}
	\psi(\rho,\theta) = \frac{\delta}{\lambda^2}\rho^2\sin^2\theta.
\end{equation}

The general solution to (\ref{eq:Bobhom}) is obtained by a separated solution $\psi(\rho,\theta) = R(\rho) \Theta(\theta)$. Upon substituting the separated form into (\ref{eq:Bobhom}) one arrives at the two ODEs

\begin{equation}\label{eq:rho_ode2}
	\rho^2 R''(\rho) - (c + \lambda^2)R(\rho) = 0
\end{equation}

\begin{equation}\label{eq:theta_ode2}
	\big((-\csc \theta) \Theta'\big)' = \mathcal{C} (\csc \theta) \Theta.
\end{equation}

One can notice that (\ref{eq:theta_ode2}) is the exact same as in the previous section given by (\ref{eq:theta_ode1}). Therefore, due to the $\sin^2\theta$ dependence in (\ref{eq:part}) and the orthogonality of $\Theta_l(\theta)$ given by (\ref{eq:thetasol}), one can conclude in a similar fashion to the previous section that the only value of $l$ which satisfies the pressure $P$ going to the constant ambient pressure P0 on the boundary is $l = 1$. This gives the following separated anzats to use

\begin{equation}
	\psi(\rho,\theta) = G(\rho)\rho^2\sin^2\theta.
\end{equation}
Upon substituting the above into equation (\ref{eq:GS_spherical1}), the second order linear ODE is obtained
\begin{equation}\label{eq:eig_ODE}
	G''(\rho) + \frac{4}{\rho} G'(\rho) + G(\rho)\lambda^2 = \gamma.
\end{equation}
This third order equation, (\ref{eq:eig_ODE}), along with the following three physical conditions gives a well posed eigenvalue problem \cite{Bobnev}.
\begin{enumerate}\label{eq:boundarycondit}
	\item
	To achieve finite energy inside the sphere $\lim_{\rho \to 0}|G(\rho)| < \infty$.
	\item
	The magnetic field components given by (\ref{eq:magn}) must vanish at the boundary for the proper asymptotic behaviour as discussed in \cite{Bobnev,kaiser1995ball}, $G'(R) = G(R) = 0$.
	\item
	The pressure must go to the constant ambient pressure $P_0$ at the boundary, $G(R) = 0$.
\end{enumerate}
A general solution to (\ref{eq:eig_ODE}) can be found to be

\begin{equation}
	G(\rho) = C_1 \frac{\rho\lambda\sin(\rho\lambda) + \cos(\rho\lambda)}{\rho^3} + C_2 \frac{\rho\lambda\cos(\rho\lambda) - \sin(\rho\lambda)}{\rho^3} + \frac{\gamma}{\lambda^2}.
\end{equation}
From the first condition above, $C_1 = 0$. The second condition gives a countable number of normalized eigenvalues $\lambda_n = \lambda R$ corresponding to the nth root of the following transcendental equation
\begin{equation}\label{eq:Bobtrans}
	x^2\tan x - 3\tan x + 3x = 0.
\end{equation}
Lastly, the third condition gives a value for $\gamma$ depending on the value of $\lambda_n$,
\begin{equation}\label{eq:Bobgamma}
	\gamma_n = -C_2\lambda_n^2\frac{\lambda_n\cos\lambda_n - \sin \lambda_n}{R^5}.
\end{equation}
This gives the flux function inside of the sphere as
\begin{equation}\label{eq:Bobnev}
	\psi(\rho,\theta) =  \left(C_2\frac{\frac{\rho}{R}\lambda_n\cos(\frac{\rho}{R}\lambda_n) - \sin(\frac{\rho}{R}\lambda_n)}{\rho} + \frac{\rho^2 R^2\gamma_n}{\lambda_n^2}\right)\sin^2\theta.
\end{equation}

which can be written in terms of a first order spherical Bessel function of the first kind, $j_1$ as

\begin{equation}\label{eq:Bobnev1}
	\psi(\rho,\theta) =  \left(\tilde{C_2}\frac{\rho}{R} \lambda_n  j_1\left(\frac{\rho}{R} \lambda_n\right) + \frac{\rho^2 R^2\gamma_n}{\lambda_n^2}\right)\sin^2\theta.
\end{equation}

\begin{figure}[htb!]	
	\begin{center}
		\includegraphics[width = .6\textwidth]{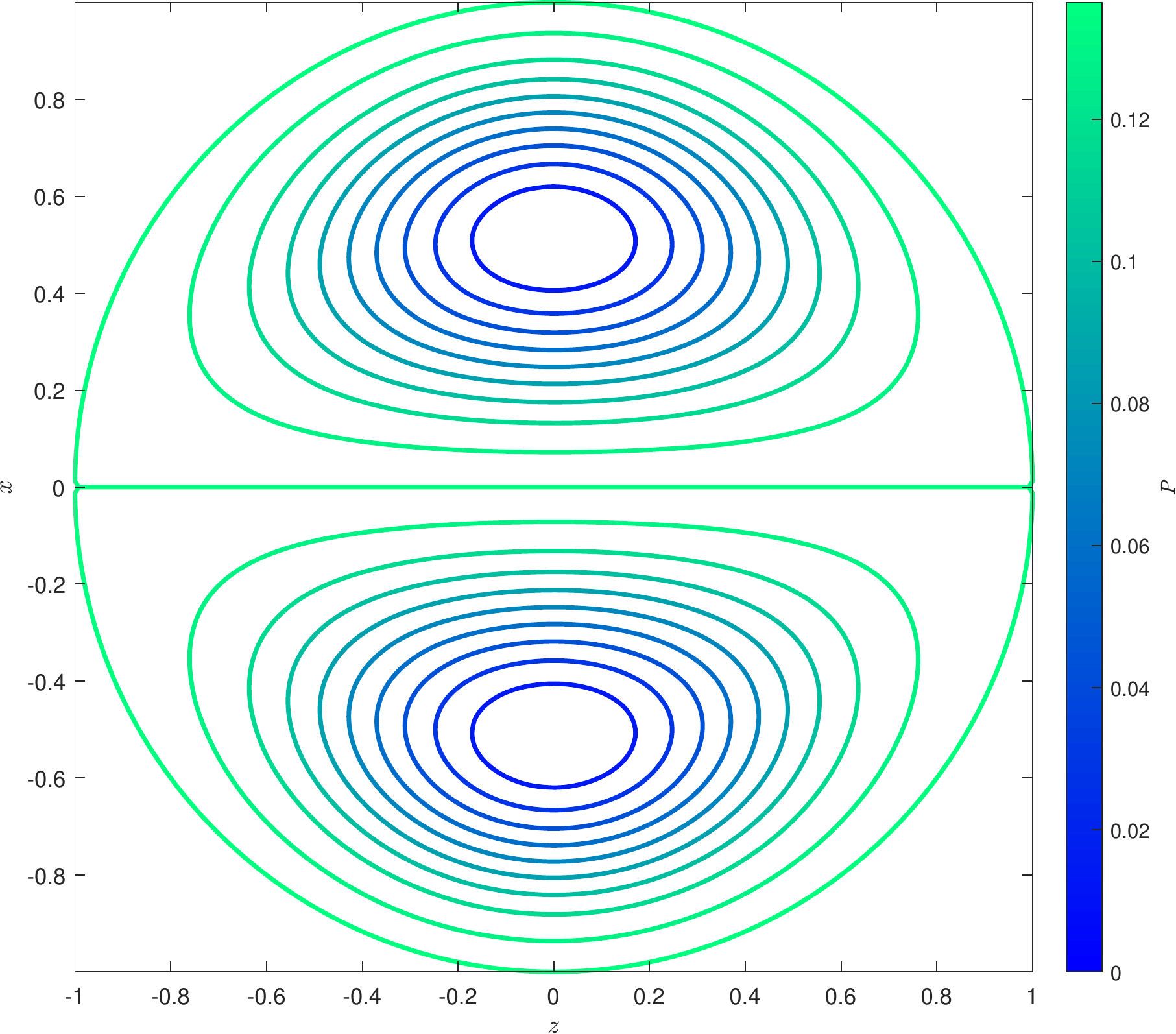}
	\end{center}
	\caption{Pressure profile of static spherical vortex in ideally conducting fluid given by $P(\psi_n) = P_0 - \gamma_n \psi_n$ where $\psi_n$ is given by (\ref{eq:Bobnev}) for $R = 1$, $n = 1$ and $C_2 = 1$. $\vec{B}$ is not shown on this plot as the non-zero $\phi$ component would make it point out of, or into the page.}\label{fig:Bobnev1}
\end{figure}

Outside of the sphere $\rho > R$ all of the magnetic field components are zero and the pressure is equal to the ambient pressure $P_0$. An example of this solution for $n = 1$ has its pressure shown in Figure \ref{fig:Bobnev1}.

\begin{figure}[htb!]
	\centering
	\begin{subfigure}[b]{0.48\textwidth}
		\centering
		\includegraphics[width=\textwidth]{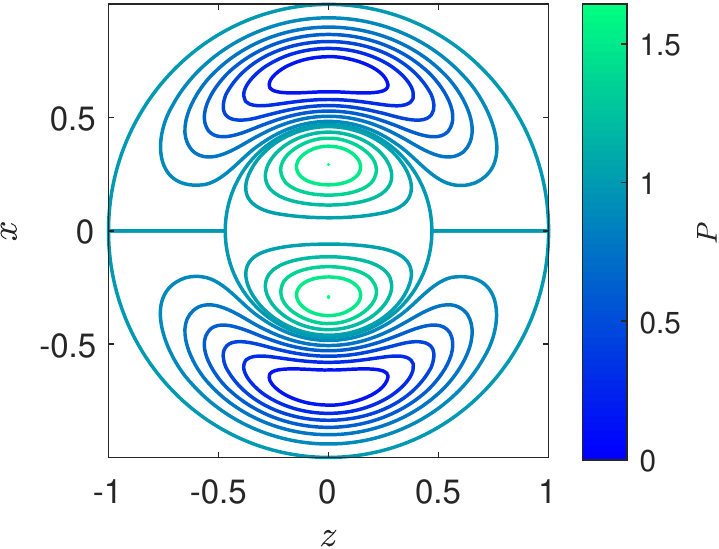}
		\caption{}\label{fig:Bobnev2}
	\end{subfigure}
	\hfill
	\begin{subfigure}[b]{0.48\textwidth}
		\hfill
		\centering
		\includegraphics[width=\textwidth]{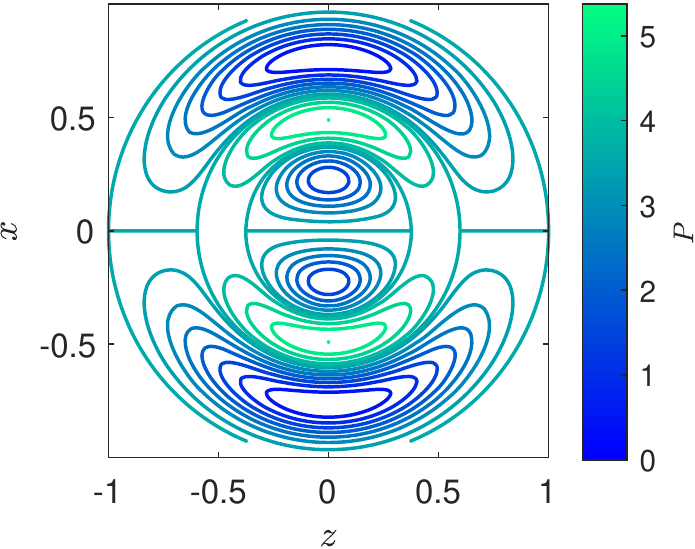}
		\caption{}\label{fig:Bobnev3}
	\end{subfigure}
	\caption{Pressure profile of static spherical vortex in ideally conducting fluid given by $P(\psi_n) = P_0 - \gamma_n \psi_n$ where $\psi_n$ is given by (\ref{eq:Bobnev}) for $C_2 = 1$, $R = 1$, $n = 2$ on the left, and $n = 3$ on the right.}\label{fig:Bobnev23}
\end{figure}

A few other solutions are shown for higher values of $n$. In Figure \ref{fig:Bobnev23} pressure profiles $P(\psi_n) = P_0 - \gamma_n\psi_n$ for $\psi_n$ given by (\ref{eq:Bobnev}) with $n = 2$ and $n = 3$ can be seen.

\begin{figure}[htb!]
	\centering
	\begin{subfigure}{0.48\textwidth}
		\centering
		\includegraphics[width=\textwidth]{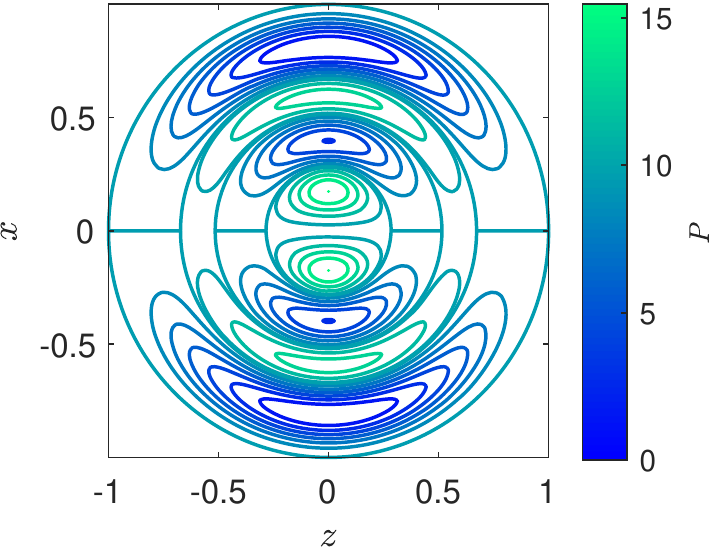}
		\caption{}\label{fig:Bobnev4}
	\end{subfigure}
	\hfill
	\begin{subfigure}{0.48\textwidth}
		\hfill
		\centering
		\includegraphics[width=\textwidth]{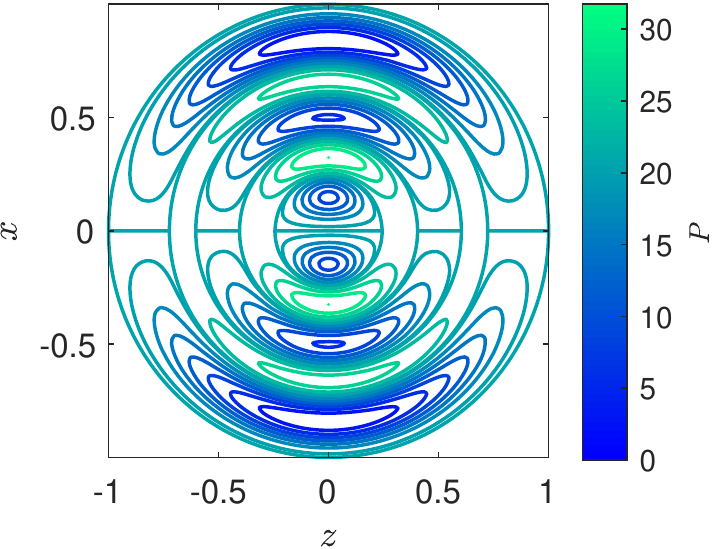}
		\caption{}\label{fig:Bobnev5}
	\end{subfigure}
	\caption{Pressure profile of static spherical vortex in ideally conducting fluid given by $P(\psi_n) = P_0 - \gamma_n \psi_n$ where $\psi_n$ is given by (\ref{eq:Bobnev}) for $C_2 = 1$, $R = 1$, $n = 4$ on the left, and $n = 5$ on the right.}\label{fig:Bobnev45}
\end{figure}

In Figure \ref{fig:Bobnev45} $n = 4$ and $n = 5$.

\section{A generalized version of Hill's spherical vortex}

In the last section, as the magnetic field outside of the spherical vortex needed to vanish in order to satisfy asymptotic behaviour that decays at least as fast as a dipole moment \cite{kaiser1995ball}, and as the velocity asymptotics of Hill's spherical vortex outside of the sphere have good behaviour from a fluid dynamics standpoint, a generalized spherical vortex with a non-zero $V^{\phi}$ can be considered in a very similar way to the previous section.
\medskip

Similar to the first section of chapter 3, using a moving frame of reference, assuming axial invariance, the Euler equations can reduce to the Bragg-Hawthorne equation. Starting from said equation in spherical coordinates

\begin{equation}\label{eq:spherical_gs}
	\left[ \frac{\partial^2}{\partial \rho^2} + \frac{\sin \theta}{\rho^2} \frac{ \partial}{\partial \theta} \left(\frac{1}{\sin \theta} \frac{\partial}{\partial \theta}\right) + F(\psi)F'(\psi)\right] \psi =  -H'(\psi)\rho^2 \sin^2{\theta},
\end{equation}
the arbitrary functions are again chosen as the highest power series expansion in $\psi$ such that the (\ref{eq:spherical_gs}) becomes separable and the asymptotics of the pressure $H(\psi)$ and the toroidal velocity component function $F(\psi)$ behave properly. As far as seperability of (\ref{eq:spherical_gs}) goes, both functions cannot be of higher degree then linear in $\psi$. In regards to the asymptotics, the pressure far away from the sphere is chosen to change and thus needs to be the ambient pressure $H_0$, similarly, $F(\psi)$ must also not change far away from the sphere, however, $F(\psi) = F_0$ where $F_0 = \const$ is not allowed as it corresponds to a singular $V^{\phi}$. This gives the best option for the free functions as

\begin{equation}
	H(\psi) =
	\begin{cases}
		H_0 - \gamma \psi, & \rho < R\\
		
		H_0 , & \rho > R
	\end{cases}
\end{equation}

\begin{equation}
	F(\psi) =
	\begin{cases}
		\lambda\psi, & \rho < R\\
		
		0. & \rho > R
	\end{cases}
\end{equation}
This allows one to decompose the spherical Grad-Shafranov equation into two problems like before, one inside and one outside of the sphere, namely:

\begin{enumerate}
	\item{Rotational flow inside the sphere $\rho < R$}
	\begin{subequations}\label{eq:11}
		\begin{equation}
			H(\psi) = H_0 - \gamma \psi,
		\end{equation}
		\begin{equation}\label{eq:Hill_inside2}
			\left[ \frac{\partial^2}{\partial \rho^2} + \frac{\sin \theta}{\rho^2} \frac{ \partial}{\partial \theta} \left(\frac{1}{\sin \theta} \frac{\partial}{\partial \theta}\right) + \lambda^2\right] \psi = \gamma \rho^2 \sin^2{\theta}.
		\end{equation}
	\end{subequations}
	
	\item{Irrotational, force-free flow outside the sphere $\rho > R$}
	\begin{subequations}\label{eq:22}
		\begin{equation}
			H(\tilde{\psi}) = H_0,
		\end{equation}
		\begin{equation}
			\left[ \frac{\partial^2}{\partial \rho^2} + \frac{\sin \theta}{\rho^2} \frac{ \partial}{\partial \theta} \left(\frac{1}{\sin \theta} \frac{\partial}{\partial \theta}\right)\right] \psi = 0.
		\end{equation}
	\end{subequations}
\end{enumerate}

The velocity components inside and outside are given by
\begin{equation}
	\vec{v}_{in} =  \frac{1}{\rho^2\sin \theta}\frac{\partial \psi}{\partial \theta} \vec{e}_{\rho} + \frac{F(\psi)}{\rho\sin\theta}\vec{e}_{\phi} - \frac{1}{\rho \sin \theta}\frac{\partial \psi}{\partial \rho} \vec{e}_{\theta},
\end{equation}
and
\begin{equation}
	\vec{v}_{out} =  \frac{1}{\rho^2\sin \theta}\frac{\partial \tilde{\psi}}{\partial \theta} \vec{e}_{\rho} - \frac{1}{\rho \sin \theta}\frac{\partial \tilde{\psi}}{\partial \rho} \vec{e}_{\theta}.
\end{equation}
respectively. Along with this, the matching pressure at the boundary, the need for $\psi(\rho,\theta)$ to be regular at $\rho = 0$ and the matching velocity at the boundary give in order the following four boundary conditions identical to the first section of Chapter 1
\begin{equation}\label{eq:matching3}
	\psi(R,\theta) = 0, \quad |\psi(0,\theta)| < \infty,  \quad \frac{\partial \psi}{\partial \theta} \bigg|_{\rho = R} = \frac{\partial \tilde{\psi}}{\partial \theta} \bigg|_{\rho = R}, \quad \frac{\partial \psi}{\partial \rho} \bigg|_{\rho = R} = \frac{\partial \tilde{\psi}}{\partial \rho} \bigg|_{\rho = R}.
\end{equation}
From the last Section, a solution inside the sphere that is bounded at the origin is found to be
\begin{equation}\label{eq:inside2}
	\psi(\rho,\theta) = \left(C\frac{\rho\lambda\cos(\rho\lambda) - \sin(\rho\lambda)}{\rho} + \frac{\gamma}{\lambda^2}\rho^2\right)\sin^2\theta,
\end{equation}
and from the first section, the solution outside of the sphere is given by
\begin{equation}
	\tilde{\psi}(\rho,\theta) =  \rho^2\sin^2\theta \left(A + \frac{B}{\rho^3}\right).
\end{equation}
\medskip

After applying the matching pressure boundary condition given by the first equation in (\ref{eq:matching3})  one obtains the transcendental equation between $\lambda$ and $\gamma$
\begin{equation}\label{eq:trans1}
	C\lambda^2R\cos(R\lambda) - C\lambda\sin(R\lambda) + R^3\gamma = 0.
\end{equation}
\medskip

Using the third boundary condition in (\ref{eq:matching3}) one obtains
\begin{equation}\label{eq:cond5}
	A = -\frac{B}{R^3}
\end{equation}

giving the outside solution as
\begin{equation}
	\tilde{\psi}(\rho,\theta) =  B\rho^2\sin^2\theta \left(\frac{1}{\rho^3} - \frac{1}{R^3}\right).
\end{equation}
\medskip

Lastly, the final boundary condition in (\ref{eq:matching3}) allows one to solve for $B$ in terms of the other constants, giving
\begin{equation}\label{eq:trans2}
	B = \frac{CR\lambda^3\cos(R\lambda)+ C\lambda^4 R^2 \sin(R\lambda) - C\lambda^2\sin(R\lambda) - 2\gamma R^3}{3\lambda^2}.
\end{equation}

The three conditions on the constants given by (\ref{eq:trans1}), (\ref{eq:cond5}) and (\ref{eq:trans2}) gives $\psi(\rho,\theta)$ in the whole space as
\begin{equation}\label{eq:general_hills}
	\psi(\rho,\theta) =
	\begin{cases}
		\left(C\frac{\rho\lambda\cos(\rho\lambda) - \sin(\rho\lambda)}{\rho} + \frac{\gamma}{\lambda^2}\rho^2\right)\sin^2\theta, & \rho < R\\
		
		\frac{CR\lambda^3\cos(R\lambda)+ C\lambda^4 R^2 \sin(R\lambda) - C\lambda^2\sin(R\lambda) - 2\gamma R^3}{3\lambda^2} \rho^2\sin^2\theta\left(\frac{1}{R^3} - \frac{1}{\rho^3}\right). & \rho > R
	\end{cases}.
\end{equation}

This solution (\ref{eq:general_hills}) of the spherical Grad-Shafranov equations (\ref{eq:11}) and (\ref{eq:22}) is a more general version of Hill's spherical vortex as:
\begin{itemize}
	\item The $\phi$ component of the velocity is non-zero inside of the sphere. Whereas Hill's original vortex solution had $V^{\phi} = 0$.
	\item There is the choice of freedom for three constants, $C$, ($\lambda$ or $\gamma$) and $R$, whereas Hill's original solution only has a choice of freedom for $R$ and one constant $\delta$.
\end{itemize}

The asymptotics of the velocity field outside of the sphere behave in a suitable manner as this is the same outside solution of Hill's spherical vortex given in cylindrical coordinates by (\ref{eq:vel1}) and (\ref{eq:vel2}) which has correct asymptotics as discussed in \cite{hill1894vi}. One interesting remark is that if the outside magnetic field must vanish which corresponds in this case to the coefficient of the outside solution given in \ref{eq:general_hills} as $B$, then this problem reduces to the problem in the previous section and the equations (\ref{eq:trans1}) and (\ref{eq:trans2}) reduce to the transcendental equations given by (\ref{eq:Bobtrans}) and (\ref{eq:Bobgamma}) as they should. This result is briefly discussed in \cite{kaiser1995ball} as they require this condition for the proper asymptotics of the magnetic field.

\section{Spherical separation of variables for Grad-Shafranov equation}

In Section 3 of this Chapter, a separated solution in spherical coordinates to the Grad-Shafranov equation \ref{eq:GS_spherical} was obtained to satisfy boundary conditions that correspond to a spherical vortex moving through a stationary fluid. During this, the behaviour of $\Theta_l(\theta)$ given by (\ref{eq:thetasol}) was restricted to $l = 1$ to satisfy the boundary conditions. In this section, a fully separated solution is considered in its own right.
\medskip

Using the first part of Section 3 up until \ref{eq:theta_ode2}, the linear Grad-Shafranov equation in spherical coordinates

\begin{equation}\label{eq:GS_spherical}
	\left[ \frac{\partial^2}{\partial \rho^2} + \frac{\sin \theta}{\rho^2} \frac{ \partial}{\partial \theta} \left(\frac{1}{\sin \theta} \frac{\partial}{\partial \theta}\right) + \lambda^2\right] \psi = \gamma \rho^2 \sin^2{\theta}.
\end{equation}
which corresponds to the free functions from Section 3 given by $I(\psi) = \lambda \psi$ and $P(\psi) = P_0 - \gamma \psi$. A solution in the form
of  $\psi(\rho,\theta) = \psi(\rho,\theta)_{gen} + \psi(\rho,\theta)_{part}$  is sought with $\psi(\rho,\theta)_{part} = \frac{\gamma \rho^2 \sin^2\theta}{\lambda^2}$. A separated solution for the homogenous version of (\ref{eq:GS_spherical}) is sought in the form $\psi(\rho,\theta) = R(\rho)\Theta(\theta)$.
\medskip

The homogeneous version of equation (\ref{eq:GS_spherical}) then reduces to the two ODEs

\begin{equation}\label{eq:rho_ode}
	\rho^2 R''(\rho) - (\mathcal{C} + \lambda^2)R(\rho) = 0,
\end{equation}

\begin{equation}\label{eq:theta_ode}
	\Theta''(\theta) - \frac{\cos\theta}{\sin\theta}\Theta'(\theta) + c\Theta(\theta) = 0,
\end{equation}
where $\mathcal{C}$ is a separation constant to be determined.
\medskip

From Section 1 of this chapter, the separation constant is found to be $\mathcal{C} = l(l+1)$ for $l \in \mathbb{N}$ with a solution to (\ref{eq:theta_ode}) given by
\begin{equation}\label{eq:theta11}
	\Theta_l(\theta) = (l+1)\mathcal{P}_{l+1}(\cos\theta) - (l+1)\cos\theta \ \mathcal{P}_l(\cos\theta).
\end{equation}
The value $\mathcal{C} = l(l+1)$ can now be substituted into (\ref{eq:rho_ode}) giving
\begin{equation}
	\rho^2 R''(\rho) - (l(l+1) + \lambda^2)R(\rho) = 0.
\end{equation}
This has a solution in terms of the Bessel function of the first kind
\begin{equation}
	R_l(\rho) = \sqrt{\rho}\mathcal{J}\left(\frac{2l+1}{2},\rho\lambda\right).
\end{equation}
So a separated solution to the homogenous version of (\ref{eq:GS_spherical}) is given by
\begin{equation}\label{eq:Sep_sphere}
	\psi_l(\rho,\theta) = \sqrt{\rho}\mathcal{J}\left(\frac{2l+1}{2},\rho\lambda\right)\Big((l+1)\mathcal{P}_{l+1}(\cos\theta) - (l+1)\cos\theta\mathcal{P}_l(\cos\theta)\Big).
\end{equation}
\medskip

As equation (\ref{eq:GS_spherical}) is linear, any linear combination of the separated solution (\ref{eq:Sep_sphere}) with the addition of the particular solution  will also be a solution. This can be written in a general way as

\begin{equation}\label{eq:GeneralSphere}
	\Psi(\rho,\theta) = \frac{\gamma \rho^2 \sin^2\theta}{\lambda^2} + \sum_{l=0}^n a_l \sqrt{\rho}\mathcal{J}\left(\frac{2l+1}{2},\rho\lambda\right)\Theta_l(\theta).
\end{equation}

\begin{figure}[htb!]	
	\begin{center}
		\includegraphics[width = .7\textwidth]{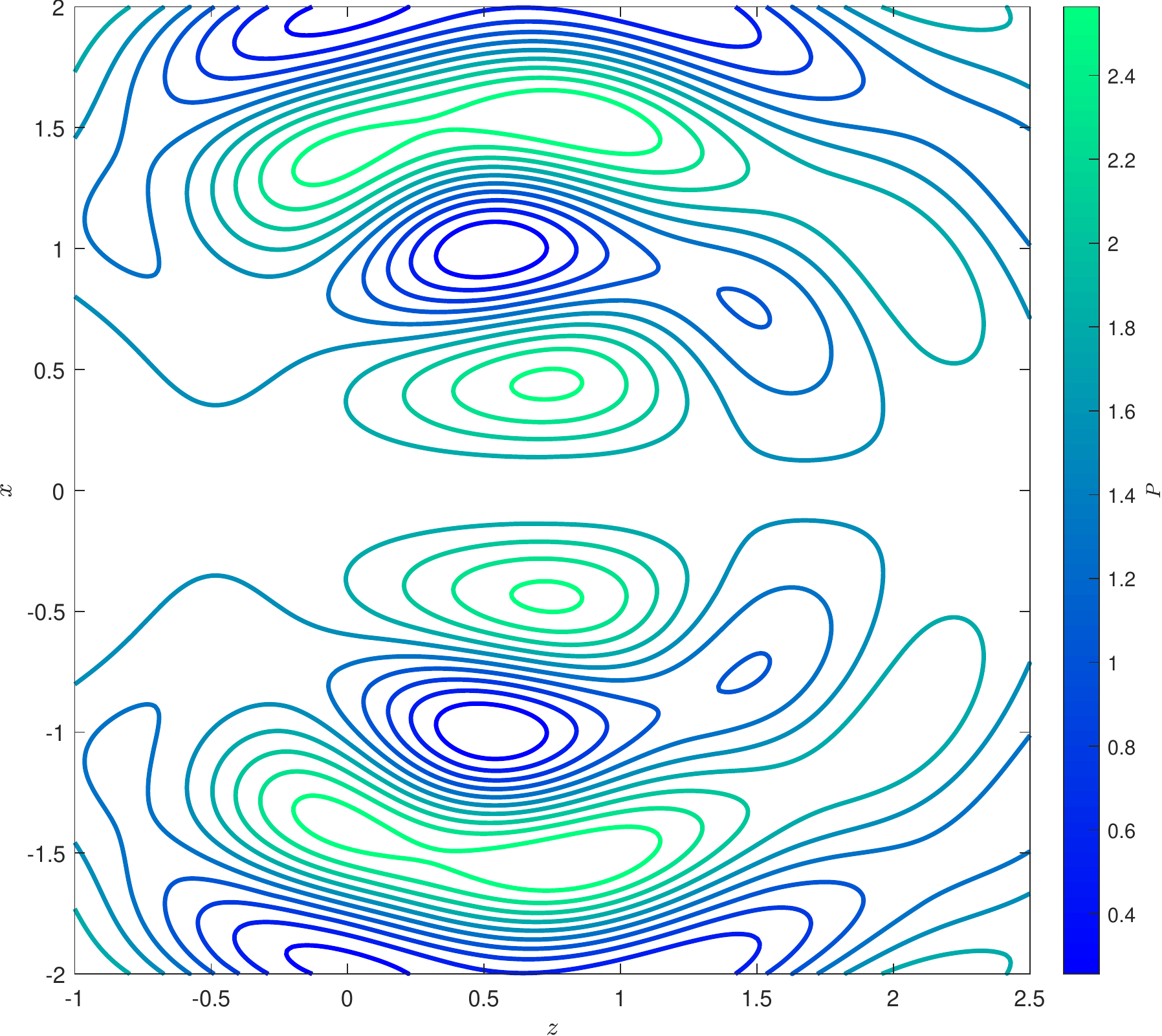}
	\end{center}
	\caption{A cross-section of magnetic surfaces where the magnetic surfaces are shown by $P(\psi) = \const$ for $P = P_0 - \gamma \psi$ where $\psi$ is given by (\ref{eq:GeneralSphere}). Here $\gamma = 1$, $\lambda = 1$, $n = 5$, $a_l = 1$, $l = 1,2,3,4,5$. Any toroidal surface can be considered a truncated solution with the outer surface described by a current sheet.}\label{fig:general_spherical}
\end{figure}

Where $\Theta_l(\theta)$ is given by (\ref{eq:theta11}). Clearly this solution is no longer related to the spherical vortex but is an MHD equilibria solution which can be considered in its own right. A pressure profile $P = P_0 - \gamma \psi$ with $\psi$ given by (\ref{eq:GeneralSphere}) can be seen in Figure \ref{fig:general_spherical}.

\section{Stability considerations for the spherical vortex}
In this section, stability of the spherical vortices solutions described in the previous chapters will be analyzed. These include Hill's vortex solutions from Section 1 given by (\ref{eq:sol1}), the MHD spherical vortex solution given in Section 2 given by (\ref{eq:Bobnev}) and the generalized Hill's vortex from Section 3 given by (\ref{eq:general_hills}). In the first part, an axially-symmetric perturbation of Hill's spherical vortex on the sphere following a method described in \cite{pozrikidis1986nonlinear} is performed with the goal of observing modes that grow exponentially in time to conclude the instability of the solution. In the next sections, a similar perturbation is attempted but is shown to not be possible. A generalized perturbation is performed with the goal of observing modes that grow exponentially in time.

\subsection{Axisymmetric perturbation of Hill's vortex}

The solution of Hill's spherical vortex at the surface of the sphere $\rho = R$ is considered. Using the dynamic equation for $\psi$ found in Hill's paper \cite{hill1894vi}

\begin{equation}\label{eq:dynamic_psi}
	\left( \frac{\partial}{\partial t} + \frac{1}{r}\frac{\partial \psi}{\partial z}  \frac{\partial}{\partial r} - \frac{1}{r}\frac{\partial \psi}{\partial r} \frac{\partial}{\partial z}\right)\left[\frac{1}{r^2}\left(\frac{\partial^2 \psi}{\partial z^2} + \frac{\partial^2 \psi}{\partial r^2} - \frac{1}{r}\frac{\partial \psi}{\partial r}\right)\right] = 0.
\end{equation}
The inside solution given by (\ref{eq:sol1}) is perturbed using
\begin{equation}\label{eq:ppert}
	\rho \mapsto \rho(1 + \epsilon h(\theta,t))
\end{equation}
giving
\begin{equation}\label{eq:pert1}
	\psi(\rho,\theta) = \delta \rho^2(1 + \epsilon h(\theta,t))^2\sin^2\theta(\rho^2(1 + \epsilon h(\theta,t))^2 - R^2).
\end{equation}
This perturbed solution is now substituted into the spherical version of the dynamic $\psi$ equation (\ref{eq:dynamic_psi}). After this, the substitution ($\rho = R$) is made and then discarding terms beyond the first order of $\epsilon$ the following third order PDE for $h(\theta,t)$ is obtained

\begin{equation}
	2R\delta\sin\theta\frac{\partial^3 h}{\partial \theta^3}  + \frac{\partial^3 h}{\partial t \partial \theta^2}  + 6R\delta\cos\theta \frac{\partial^2  h}{\partial \theta^2} + 3\frac{\cos\theta}{\sin\theta} \frac{\partial^2 h}{\partial t \partial \theta} - \frac{40R\delta}{\sin\theta} \left(\cos^2\theta - \frac{17}{20}\right)\frac{\partial h}{\partial \theta} + 20\frac{\partial h}{\partial t} = 0.
\end{equation}

This linear homogeneous equation is separable: one can seek its solutions as $h(\theta,t) = \Theta(\theta)T(t)$ where $\Theta(\theta)$ and $T(t)$ satisfy

\begin{equation}\label{eq:Theta}
	\frac{d^3 \Theta}{d \theta^3} = -3\left(\frac{\cos\theta}{\sin\theta} + \frac{\lambda}{6R\delta\sin\theta}\right)\frac{d^2\Theta}{d \theta^2} + \left(20\frac{\cos^2\theta}{\sin^2\theta} - 3\frac{\cos\theta}{2R\delta\sin^2\theta}\lambda - \frac{17}{\sin^2\theta}\right)\frac{d\Theta}{d\theta} - 10\frac{\lambda}{R\delta\sin\theta}\Theta,
\end{equation}
\begin{equation}
	\frac{dT}{dt} = \lambda T.
\end{equation}

The $T$ equation above has the exponential solution $T(t) = Ae^{\lambda t}$. The $\Theta$ equation (\ref{eq:Theta}) can be converted into a simpler equation with the transformation $z = \cos\theta$ with $\Theta(\theta) = Z(z)$. This gives

\begin{equation}\label{Degen_ode}
	(1 - z^2)\frac{d^3 Z}{dz^3} - \left(2K_2 + 6z\right)\frac{d^2 Z}{dz^2} + 8\left(2 + \frac{K_2z}{1-z^2}\right)\frac{dZ}{dz} - \frac{40K_2}{1-z^2}Z = 0.
\end{equation}

Solutions to (\ref{Degen_ode}) can be expressed as a linear combination of the following functions written in terms of the hypergeometric functions

\begin{subequations}\label{eq:Zss}
	\begin{equation}\label{eq:Z1}
		Z_1 = \mathcal{H}\left(\left[\frac{3}{4} + \frac{\sqrt{89}}{4} , \frac{3}{4} - \frac{\sqrt{89}}{4}\right],\frac{1}{2},z^2\right),
	\end{equation}
	\begin{equation}
		Z_2 = z\mathcal{H}\left(\left[\frac{5}{4} + \frac{\sqrt{89}}{4} , \frac{5}{4} - \frac{\sqrt{89}}{4}\right],\frac{3}{2},z^2\right),
	\end{equation}
	\begin{equation}
		Z_3 = -Z_1\int_{z_0}^{z} z(z + 1)^{1 + \frac{\lambda}{4R\delta}}(z - 1)^{1 -\frac{\lambda}{4R\delta}} Z_2 dz + Z_2\int_{z_0}^{z} (z + 1)^{1 + \frac{\lambda}{4R\delta}}(z - 1)^{1 -\frac{\lambda}{4R\delta}} Z_1 dz.
	\end{equation}
\end{subequations}
Here $z_0$ is any constant such that $z_0 < z$. One should notice that both the first and second solution of (\ref{eq:Zss}) do not depend on the separation constant $\lambda$. This is because (\ref{Degen_ode}) can be written as
\begin{equation}
	\mathcal{L} = \left(\frac{d}{dz}  - \frac{2K_2}{1-z^2}\right) \mathcal{G},
\end{equation}
where
\begin{equation}\label{Degen_ode_aux}
	\mathcal{G} \equiv \left(1-z^2\right)\frac{d^2 Z}{dz^2} - 4z\frac{d Z}{dz} + 20 Z = 0.
\end{equation}
Here (\ref{Degen_ode_aux}) has the general solution
\begin{equation}
	Z = C_1 Z_1 + C_2 Z_2
\end{equation}
where $Z_1$ and $Z_2$ are given in (\ref{eq:Zss}).

\begin{figure}[htb!]	
	\begin{center}
		\includegraphics[width = .45\textwidth]{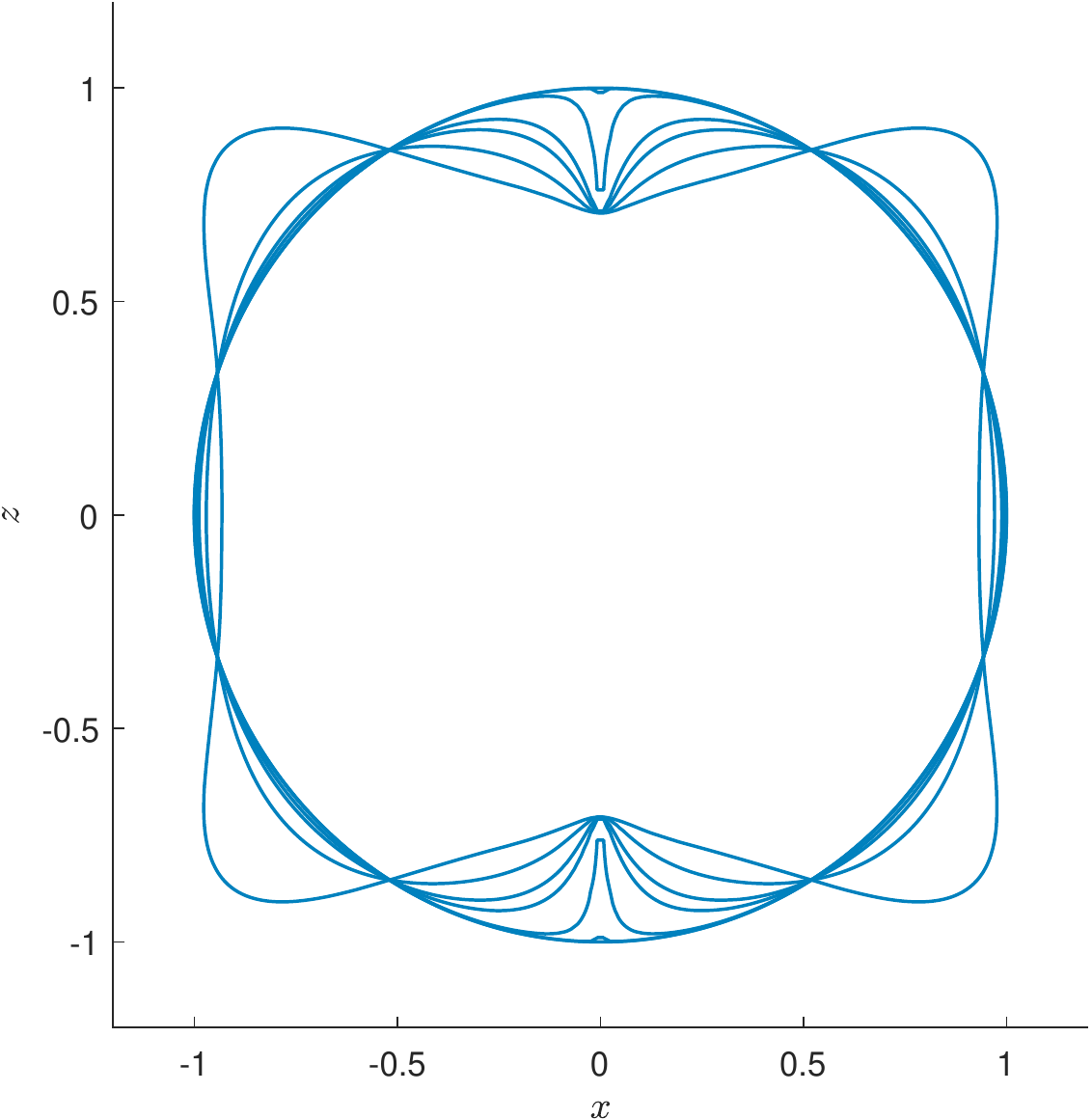}
	\end{center}
	\caption{The evolution of the perturbation given by \ref{eq:psipert2} is shown for $\epsilon = 0.0001$, $\delta =  1$, $R = 1$, $A = 1$, $\lambda = 1$, at several different times $0<t<16$. These surfaces are regular.}\label{fig:goodpert}
\end{figure}

As $\lambda$ does not appear in $Z_1$ and $Z_2$, there will exist $h(\theta,t)$ which grows exponentially in time as $\lambda$ can by positive. However, one must check and make sure that these $h(\theta,t)$ that grow in time correspond to regular surfaces. One such $h(\theta,t)$ that gives regular surfaces utilizes $Z_1$ given above by \ref{eq:Z1}. This gives $h(\theta,t)$ as

\begin{equation}
	h(\theta,t) = Ae^{\lambda t}\mathcal{H}\left(\left[\frac{3}{4} + \frac{\sqrt{89}}{4} , \frac{3}{4} - \frac{\sqrt{89}}{4}\right],\frac{1}{2},\cos^2\theta\right)
\end{equation}
This is now substituted into (\ref{eq:pert1}). After expanding out, and converting back to cylindrical coordinates, one arrives at

\begin{equation}\label{eq:psipert2}
	\psi(r,z,t) = -2\delta \left(Ae^{\lambda t}\epsilon^2(R^2 - 2r^2 - 2z^2)\mathcal{H}\left(\left[\frac{3}{4} + \frac{\sqrt{89}}{4} , \frac{3}{4} - \frac{\sqrt{89}}{4}\right],\frac{1}{2},\frac{r^2}{r^2 + z^2}\right) + \frac{R^2 - r^2 - z^2}{2}\right).
\end{equation}
When $\psi = 0$ this corresponds to the boundary of the sphere. Several plots of the evolution of this surface are shown in \ref{fig:goodpert}.

Despite the irregular look of this surface at some of the points in time, the implicit derivative ${dz}/{dr}$ of (\ref{eq:psipert2}) when $\psi(r,z,t) = 0$ can be shown to be zero at the irregular looking points $r = 0$.
\medskip

The above analysis leads to the following conclusion that Hill's spherical vortex is in general not linearly stable with respect to magnetic surface perturbations described by (\ref{eq:pert1}). Stability analysis of Hill's vortex has been previously considered numerically in ref \cite{pozrikidis1986nonlinear}, however, no details including mathematical formulas, numerical method used, and initial/boundary conditions were presented; we were not able to reproduce the results of \cite{pozrikidis1986nonlinear}.

\subsection{An axisymmetric perturbation of generalized Hill's spherical vortex and MHD vortex}

The goal here is to use a similar axisymmetric perturbation method as above following the method in \cite{pozrikidis1986nonlinear} to study the stability of the generalized hill's spherical vortex solution (\ref{eq:general_hills}) and the MHD spherical vortex in an ideally conducting fluid solution (\ref{eq:Bobnev}). The dynamic equation for $\psi$ (\ref{eq:dynamic_psi}) taken from \cite{hill1894vi} in which $V^{\phi} = 0$ was used to study the time evolution of $\psi$ with the perturbation given by equation (\ref{eq:ppert}). Therefore for a similar analysis of the two other solutions, a dynamic equation for $\psi$ needs to be derived from the time dependent, axially symmetric Euler equations with $V^{\phi} = I(\psi)/r$ (which is the form of $V^{\phi}$ in both (\ref{eq:Bobnev}) and (\ref{eq:general_hills})).

\subsubsection{Deriving axially symmetric dynamic $\psi$ equation with non-zero $V^{\phi}$}

Starting with the dynamic Euler Equations (\ref{eq:Euler3}) in cylindrical coordinates with axial invariance one arrives at the system

\begin{subequations}\label{eq:Axial_Euler}
	\begin{equation}
		V^r_t + rV^z(V^r_z -V^z_r) - V^\phi(rV^\phi)_r = rH_r,
	\end{equation}
	\begin{equation}\label{phi_equation}
		V^{\phi}_t + V^z(rV^\phi)_z + V^r(rV^\phi)_r = 0,
	\end{equation}
	\begin{equation}
		V^z_t + V^r(V^z_r - V^r_z) - V^\phi V^\phi_z = H_z,
	\end{equation}
	\begin{equation}
		(rV^z)_z + (rV^r)_r = 0,
	\end{equation}
\end{subequations}
where superscripts denote the vector component and subscripts denote the partial differentiation. The last equation, by the Pointcar\'e lemma, implies the local existence of a potential such that

\begin{equation}
	V^r = \frac{\psi_z}{r}, \quad
	V^z = -\frac{\psi_r}{r}.
\end{equation}
\medskip

Upon substituting the above vector components and the form of $\phi$ component of the velocity to be $V^{\phi} = I(\psi)/r$, (as taken from both the MHD spherical vortex solution and generalized hill solution) into (\ref{phi_equation}) one obtains

\begin{equation}\label{phi_equation2}
	I'(\psi)\frac{\partial \psi}{\partial t} = 0.
\end{equation}

This implies that for $V^{\phi} = I(\psi)/r$, either

\begin{enumerate}
	\item
	$\psi(r,z,t)$ is time independent, in which case (\ref{eq:Axial_Euler}) can reduce to the Grad-Shafranov (Bragg Hawthorn) equation.
	\item
	$I(\psi)$ is constant with respect to $\psi$. For the case when $I(\psi) = 0$, \ref{eq:dynamic_psi} can be obtained.
\end{enumerate}

Therefore, either $\psi$ is time independent or $V^{\phi} = I(\psi)/r$ is not the correct form of $V^{\phi}$ concluding that no dynamic $\psi$ equation with  $V^{\phi} = I(\psi)/r$ can exist. If $V^{\phi}$ is an arbitrary function of $r$, $z$ and $t$, with the use of Poisson Brackets, dynamic equations for $\psi$ were derived in \cite{bogoyavlenskij2010restricted}. However, these are of no use for studying the case when $V^{\phi} = I(\psi)/r$. Therefore, the time evolution of $\psi$ using a single equation is not possible and more general type of perturbation analysis needs to be considered.

\subsection{A general linear perturbation for generalized Hill's spherical Vortex}

In this section, finding solutions to the general linear perturbations was not successful, however the following methodology is still presented to show how one can derive the perturbed linear systems.
\medskip

In order to study the stability of the solution given by (\ref{eq:general_hills}) a linear perturbation on the dependent variables will be considered. As the $r$ and $z$ components of $\vec{V}$ are related by the stream function $\psi$ by
\begin{equation}\label{vel_comp}
	\vec{v} =  \frac{\psi_z}{r} \vec{e}_r  + \frac{-\psi_r}{r} \vec{e}_z,
\end{equation}
and the other dependent variables being the pressure $H$ and the $\phi$ component of the magnetic field $V^{\phi}$, instead of the usual four dependent variables, there are only three. These three quantities are perturbed as follows

\begin{subequations}\label{eq:Euler_perturbation}
	\begin{equation}
		\psi(r,z,t) = \psi_0(r,z) + \epsilon\psi_1(r,z,t),
	\end{equation}
	\begin{equation}
		F(r,z,t) =  F_0 (\psi_0) + \epsilon F_1(r,z,t),
	\end{equation}
	\begin{equation}
		H(r,z,t) = H_0(\psi_0) + \epsilon H_1(r,z,t).
	\end{equation}
\end{subequations}
where $\psi_0$ is the static solution given by (\ref{eq:general_hills}), $F_0(\psi_0) = \lambda \psi_0$ and $H_0(\psi_0) = H_0 - \gamma \psi_0$. Substituting these into the axially invariant Euler equations and discarding terms of $\epsilon^2$ and higher one obtains a closed linear system for the three unknown functions $\psi_1(r,z,t)$, $F_1(r,z,t)$, $H_1(r,z,t)$. The goal now is to see if any solutions to this linear system have time dependence that grows unbounded. The system, though linear is still very large and complex (so much so that it is not even written here), and no meaningful nontrivial solutions were able to be found to this variable coefficient linear system.

\subsection{General perturbation for an MHD spherical vortex}

Similarly to above, one can considered the perturbation of the MHD spherical vortex with the solution given by  (\ref{eq:Bobnev}). The main difference from the previous section being that the magnetic field components are perturbed as well as the velocity field components. The static equilibrium MHD equations, $\div{\vec{B}} = 0$ gives the condition that $B^r = \psi_z/r$ and $B^z = -\psi_r/r$. Also from $\div{\vec{V}} = 0$ gives the condition that $V^r = \xi_z/r$ and $V^z = -\xi_r/r$ This gives 5 dependent variables $\psi(r,z,t)$, $\xi(r,z,t)$, $I(r,z,t)$, $F(r,z,t)$ and $P(r,z,t)$, instead of the usual 6. These quanities are perturbed and written as

\begin{subequations}
	\begin{equation}
		\psi(r,z,t) = \psi_0(r,z) + \epsilon\psi_1(r,z,t),
	\end{equation}
	\begin{equation}
		I(r,z,t) =  I_0 (\psi_0) + \epsilon I_1(r,z,t),
	\end{equation}
	\begin{equation}
		P(r,z,t) = P_0(\psi_0) + \epsilon P_1(r,z,t).
	\end{equation}
	\begin{equation}
		\xi(r,z,t) = 0 + \epsilon\xi_1(r,z,t),
	\end{equation}
	\begin{equation}
		F(r,z,t) =  0 + \epsilon F_1(r,z,t),
	\end{equation}
\end{subequations}
where $\psi_0$ is given by \ref{eq:Bobnev}. Here $B^{\phi} = I(r,z,t)/r$ and $V^{\phi} = F(r,z,t)/r$.
Substituting these into the axially MHD equations gives an overdetermined system of 6 equations for the 5 unknown $\psi_1(r,z,t)$, $I_1(r,z,t)$, $P_1(r,z,t)$, $\xi_1(r,z,t)$ $F_1(r,z,t)$, and $H_1(r,z,t)$. Similar to above, no solutions were able to be found to this variable coefficient linear system.

\subsubsection*{Acknowledgements}

The authors are grateful to NSERC of Canada for the financial support
%

{\footnotesize
\bibliography{bibsph01}
\bibliographystyle{ieeetr}
}

\end{document}